\documentclass[a4paper,11pt]{article}
\usepackage{moreverb}
\usepackage{graphicx,amssymb,amstext,amsmath,tikz}
\usepackage[para]{threeparttable}
\usepackage{mathtools,xparse}
\usepackage{natbib}
\usepackage{bbold}

\usetikzlibrary{positioning,shapes.geometric}
\usepackage{rotating}
\usepackage[T1]{fontenc}
\usepackage{longtable}

\usepackage{float}
\usepackage[caption = false]{subfig}
\usepackage{natbib}

\usepackage[left=2cm,top=2cm,right=2cm, bottom=2.5cm]{geometry}
\usepackage{setspace}
\DeclarePairedDelimiter{\norm}{\lVert}{\rVert}

\def\R{\mathbb R}
\def\mM{\mathcal M} 
 
\def\mP{\mathcal{P}} 
\def\mS{\mathcal{S}}

\def\mao{\pi_{0}} 

\def\ci{\perp\!\!\!\perp}

\date{} 
\begin{document}
\title{Data-adaptive doubly robust instrumental variable methods for treatment effect heterogeneity}

\date{}
\author{K.~DiazOrdaz,  R.~Daniel, N. Kr\'eif  \\Department of Medical Statistics, LSHTM,\\
Division of Population Medicine,
Cardiff University, \\Centre for Health Economics, University of York.\\
}


\maketitle

\abstract{
We consider the estimation of the average treatment effect in the treated as a function of baseline covariates, where there is a valid (conditional) instrument.

We describe two doubly robust (DR) estimators: a locally efficient g-estimator, and a targeted minimum loss-based estimator (TMLE). These two DR estimators can be viewed as generalisations of the two-stage least squares (TSLS) method to semi-parametric models that make weaker assumptions. We exploit recent theoretical results that extend to the g-estimator the use of data-adaptive fits for the nuisance parameters. 

A simulation study is used to compare standard TSLS with the two DR estimators' finite-sample performance, (1) when fitted using parametric nuisance models, and (2) using data-adaptive nuisance fits, obtained from the Super Learner, an ensemble machine learning method.

Data-adaptive DR estimators have lower bias and improved coverage, when compared to incorrectly specified parametric DR estimators and TSLS. When the parametric model for the treatment effect curve is correctly specified, the g-estimator outperforms all others, but when this model is misspecified, TMLE performs best, while TSLS can result in large biases and zero coverage. 

Finally, we illustrate the methods by reanalysing the COPERS (COping with persistent Pain, Effectiveness Research in Self-management) trial to make inferences about the causal effect of treatment actually received, and the extent to which this is modified by depression at baseline.}

\maketitle

\section{Introduction}
There has been an increased interest in estimating the causal effect of treatment actually received  in randomised controlled trials (RCTs) in the presence of treatment non-adherence, in addition to the intention-to-treat effect,  as highlighted by the International Council for Harmonisation addendum to guideline E9 (Statistical Principles for Clinical Trials, addendum on Estimands).   An additional challenge is posed by appreciable treatment effect heterogeneity, which is often itself of interest. This is a a common issue with psychological interventions \citep{Dunn2007}.  

In this work, we consider methods for estimating the dependence of a causal average treatment effect on baseline covariates in RCTs with non-adherence. This is motivated by the COPERS (COping with persistent Pain, Effectiveness Research in Self-management) trial. The intervention introduced cognitive behavioural therapy approaches designed to promote self-efficacy to manage chronic pain, with the primary outcome being pain-related disability. The research team  was interested in the causal effect  of the received treatment, and whether this effect was modified by a number of baseline variables.  Here, we will focus on  one possible effect modifier: depression at baseline, measured by the  Hospital Anxiety and Depression Scale (HADS).

Instrumental variable (IV) methods are often used to estimate the effect of treatment received in RCTs where randomised treatment is unconfounded by design, but treatment received is not.
  Assuming that randomised treatment is a valid instrument, and under some additional assumptions reviewed in Section \ref{Sec:IV}, it is possible to identify the average treatment effect in the treated, conditional on baseline covariates $V$. In addition to investigating effect modification by a subset of baseline covariates $V$, it can be beneficial to use a larger set $W$ of baseline covariates for adjustment in the analysis:  (i) if  the IV assumptions  are more plausible conditional on baseline covariates $W$, or  (ii) to increase the statistical efficiency of the estimators.

A relatively simple  method  of estimation for this is the so-called  two stage least squares (TSLS). In its simplest form, i.e. when $V$ is null, the first stage predicts the exposure  based on an ordinary least squares regression of the exposure on the IV and baseline covariates $W$, while the second stage regresses the outcome on the predicted exposure  from the first stage  and baseline covariates $W$, also via ordinary least squares regression. The coefficient corresponding to the predicted exposure in this second model is  the TSLS estimator of the desired causal treatment effect. TSLS is robust to the misspecification  of the first stage model \citep{Robins2000, Wooldridge2010} but may be inefficient,  especially when the treatment-exposure relationship is non-linear  \citep{Vansteelandt2015}. However,  where $V$ is non-null and the treatment effect varies by baseline covariates, TSLS relies on the outcome model (the second stage) being correctly specified to obtain consistent effect estimates.

 Doubly robust (DR) estimators are appealing in such settings, as they  estimate  consistently the  parameter of interest  if at least one of the models,  for either the exposure  or the outcome is correctly specified.   In the context of linear IV models with $V$ null,   \citet{Okui2012} proposed  a locally-efficient estimating equations DR estimator for the causal effect of treatment in the treated, often called a \textit{g-estimator}. It augments the TSLS estimating equation by adding a model for the instrument given the baseline covariates. This estimator is DR in the sense that it needs to specify correctly either the  outcome model or the instrument model.  This estimator was generalised to settings where $V$ is non-null by \citet{Vansteelandt2015} and shown to be locally efficient.
  
Recently, \citet{Toth2016}  proposed a DR targeted maximum likelihood estimator (TMLE) for the treatment effect in a  linear IV model. TMLE is a general approach for causal inference problems yielding  semi-parametric substitution estimators \citep{van2011targetedbook}. 

Although DR estimators offer in principle partial protection against model misspecification,  concerns remain over their performance in practice, when all models are likely to be misspecified \citep{Kang2007}. To alleviate biases due to model misspecification, TMLE is usually coupled with machine learning estimation of the nuisance parameters, using in particular the Super Learner, a cross-validation based estimator selection approach \citep{van2007super}.  TMLE and  other DR estimators possess a particular orthogonality property that leads to greater suitability with machine learning estimation. Estimators based on a single nuisance model can perform poorly when combined with machine learning fits, since the estimator inherits the slow convergence (and hence high finite sample bias) and non-regularity of the machine learning estimators, with the latter phenomenon making valid statistical inferences  complex to obtain \citep{VanderVaart2014}. In addition, since the resulting estimators can be irregular, the nonparametric bootstrap is in general not  valid \citep{Bickel1997}.
Some DR estimators, such as TMLE, on the other hand, when combined with machine learning estimation of the nuisance functionals, have faster convergence and make (asymptotic) analytic statistical inference tractable via the sampling variance of the corresponding influence functions, under empirical processes conditions, assuming that the convergence rates of the machine learning estimators (to their respective truths) used are fast enough \citep{van2006targeted,Farrell2015}.  

Building on previous literature that establishes conditions for one-step and estimating equations estimators to be  (asymptotically) Neyman orthogonal \citep{Newey1994, LaanRobins2003} as well as previous work that used sample splitting to avoid empirical processes conditions
\citep{Bickel1982}, \citet{Chernozhukov2016}  proposed the use of sample splitting  when using  machine learning  for estimating the nuisance parameters, thus widening the class of estimating equations DR estimators that can be estimated data-adaptively. In particular,  \citet{Chernozhukov2016} give regularity conditions for estimating equations estimators of the linear IV model,  which can be adapted for the g-estimator introduced by \citet{Vansteelandt2015}.  
Thus, we implement here the  g-estimator  with and without machine learning estimation for nuisance parameters.  We compare its performance with that of a TMLE \citep{Toth2016}, again implemented either parametrically or data-adaptively, and standard parametric TSLS, in terms of mean bias,  root mean squared error (RMSE) and confidence interval (CI) coverage using a simulation study. We also contrast the methods by applying them to the illustrative RCT.

This paper is organised as follows. In the next section,  we define the causal parameters of interest and the assumptions for the IV methods. In Section \ref{Sec:TSLS} we  review the standard TSLS, while in Section \ref{Sec:gest}, we introduce the g-estimator  proposed by \citet{Vansteelandt2015}. Section \ref{Sec:dataadapt} briefly justifies the use of machine learning estimation for the nuisance models of the DR estimators, and introduces the Super Learner. The  TMLE estimator  proposed by  \citet{Toth2016} is described in Section \ref{Sec:TMLE}. In Section \ref{Sec:Simulation}, we present a simulation study, comparing the performance of these estimators.  The proposed methods are then applied to the COPERS RCT in Section \ref{Sec:COPERS}.  We conclude with a discussion in Section \ref{Sec:discussion}.

\section{Linear instrumental variables models}\label{Sec:IV}

 \begin{figure}\caption{
DAG depicting   a valid  conditional instrument $Z$ for exposure $A$ in the presence of observed  and unobserved confounders $W$ and $U$ respectively, where the outcome is  $Y$.}\label{dag}
\begin{small}
\begin{center}
\begin{tikzpicture}[%
->,
shorten >=2pt,
>=stealth,
node distance=1cm,
pil/.style={
->,
thick,
shorten =2pt,}
]
\node (1) {$W$};
\node[below right=of 1] (2) {$Z$};
\node[right=of 2] (3) {$A$};
\node [circle, draw] (4) [below right=of 3] {U};
\node  (5) [right=of 3] {$Y$};
\draw [->] (1) -- (2);
\draw [->] (1) -- (3);
\draw [->] (1) -- (5);
\draw [->] (2) -- (3);
\draw [->] (3) -- (5);
\draw  (4) edge[bend left=50,->] (1);
\draw [->] (4) -- (3);
\draw [->] (4) -- (5);

\end{tikzpicture}
\end{center}

\end{small}
\end{figure}
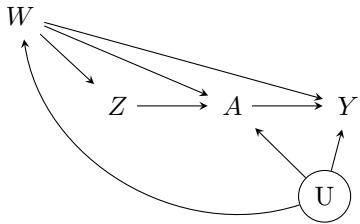

Let   $W$ be a set of baseline variables, $Z$ be the randomised treatment indicator and $A$ be the exposure of interest, the actual treatment received, assumed to be  binary. Denote by $Y$ the continuous outcome of interest, and by $U$ the set of all unobserved common causes of $A$ and $Y$.  Further,  assume that $(U, W)$ would be a sufficient set to control for the confounding in the effect
of $A$ on $Y$, were $U$ observed.  For simplicity, but without loss of generality,   we assume that interest lies in estimating effect modification by a single baseline variable $V\in W$. 

Let a subscript 0 denote the true probability distributions,  models and parameters. Let  the  vector of the observed data for the $i$-th individual be
 $O_i=\{W_i, Z_i, A_i, Y_i\}\sim P_0$, where $P_0$ is the true underlying distribution from which an independent identically distributed random sample of size $n$  is  drawn.  The causal relationships between these variables are encoded by the directed acyclic graph (DAG) shown in Figure \ref{dag}. 

Let the potential outcome $Y(a)$ be the outcome that would occur if $A$ were set to $a\in\{0,1\}$. As usual, we  assume
 \textit{no interference}, i.e. the potential outcomes of the $i$-th individual are unrelated to the treatment status of all other individuals, and \textit{counterfactual consistency},  for all individuals  $Y = Y(z)$ and $A = A(z)$ if $Z = z$, and $Y = Y(z,a)$ if $(Z,A) = (z, a)$, for all $z$ and all $a$ \citep{Rubin1978causal,VanderWeele2009consistency}. 

Following \citet{Abadie2003} and  \citet{Vansteelandt2015}, we write the conditional version of the IV assumptions  \citep{Angrist1996}, as follows:\\
 \noindent\textbf{(i) Conditional unconfoundedness}: $Z$ is conditionally  independent of  the unmeasured confounders, conditional on measured covariates $W$, i.e. $Z\ci U \vert W$.   \\  
\textbf{(ii) Exclusion restriction}: Conditionally on  $W$, $A$ and the confounder $U$, the instrument $Z$ and the response $Y$ are independent, i.e. $Z \ci Y \vert W,U, A$,
\\ \textbf{(iii) Instrument relevance} (also referred to as first stage assumption): $Z$ is associated with $A$ conditional on $W$, i.e. $Z \not\perp\!\!\!\perp A \vert W$. 

Assumptions (i) and (ii) can be shown to imply (ii') $Y(a)\ci  Z | W$, for all $a$, which is an alternative assumption often invoked independently \citep{Robins1994,Vansteelandt2015, Swanson2018}.

In addition to these IV assumptions, we assume the following  partially linear conditional mean  model for the outcome:
\begin{equation}\label{struct_modY}
  E[Y\vert A,W,Z, U]=   \varpi_0(W, U) + A m_0(W),
\end{equation} 
where   $\varpi_0(W,U)$ and $m_0(W)$  are  unknown functions, with $m_0(W)$ representing the causal treatment effect curve given covariates $W$. The assumption of linearity in $A$ is necessary to identify $m_0(W)$ using an instrument. With  binary exposure $A$, this assumption always holds.

Under these assumptions, the conditional mean model  \eqref{struct_modY} implies the so-called linear structural mean model \citep{Robins1994}:
\begin{equation}\label{IV-g1}
E[Y\vert A,W,Z]= E[Y(0)\vert A,W,Z] + A m_0(W).
\end{equation}
\textit{Proof:}
We begin by observing that eq. \eqref{struct_modY} implies $ E[Y\vert A=0,W,Z, U]=   \varpi_0(W, U).$
Thus, we can re-write eq. \eqref{struct_modY} as
\begin{eqnarray*}
 A m_0(W) &=&  E[Y\vert A,W,Z, U]-  E[Y\vert A=0,W,Z, U],\\
&=& E[Y\vert A,W,Z, U]-  E[Y(0)\vert A=0,W,Z, U],
\end{eqnarray*}
where we use the fact that $Y=Y(z,a)=Y(a)$ by counterfactual consistency and exclusion restriction. Now, since $Y(a)\ci A \vert (U, W)$, since $U$ and $W$ are sufficient to control for confounding between $A$ and $Y$, we have $E[Y(0)\vert A=0,W,Z, U]= E[Y(0)\vert A,W,Z, U]$, and thus
\begin{eqnarray*}
 A m_0(W) &=& E[Y\vert A,W,Z, U]-  E[Y(0)\vert A,W,Z, U]\\
&=& E[Y\vert A,W,Z]-  E[Y(0)\vert A,W,Z], 
\end{eqnarray*}
where the last step uses the fact that the right hand side, $Am_0(W)$ is independent of $U$ \citep{Vansteelandt2015} $\blacksquare$.

While the linear structural mean model eq. \eqref{IV-g1} can be motivated from model \eqref{struct_modY}, it is often used explicitly as the departure point for causal treatment effect estimation.   In fact, \citet{Vansteelandt2015} show that these two IV  models  imply the same restrictions on the observed data distribution, namely $E[Y-Am_0(W) \vert Z, W] = E[Y - Am_0(W) | W]$. Therefore, we denote by $\mathcal{M}$  the statistical model for $P_0$ implied by the IV assumptions  and either model \eqref{struct_modY} or \eqref{IV-g1}. This is often called  the  linear IV model.
Note that model  $\mathcal{M}$ assumes the treatment effect curve $ m_0(W)$ does not depend on $Z$. This is known as the `no effect modification' by $Z$ assumption  \citep{Hernan2006}. 

The  causal effect of interest, the average treatment effect in the treated, conditional on $V\in W$ taking the value $v$, can be written as  a function of $v$ as
\begin{equation}
\mbox{ATT}(v)=E[Y(1)-Y(0)\mid A=1, V=v]=E[m_0(W)\vert  A=1, V=v].
\end{equation}
Since  $\mbox{ATT}(v)$ is the conditional expectation of $m_0(W)$ given $A=1$ and $V=v$,  we  focus on identifying  $m_0(W)$.

Rearranging equation \eqref{IV-g1}, we have 
\begin{eqnarray}
\nonumber E[Y\vert A,W,Z]- A m_0(W) &=& E[Y(0)\vert A,W,Z],  \\
\nonumber E\left\{E[Y\vert A,W,Z]\vert W,Z\right\} - E[A m_0(W) \vert W,Z]&=& E\left\{E[Y(0)\vert A,W,Z]  \vert W,Z\right\},\\
\nonumber E[Y- Am_0(W) \vert Z, W]&=& E[Y(0)\vert Z,W], \\
E[Y- Am_0(W) \vert W] &=& E[Y(0)\vert W], 
\end{eqnarray} where in the second step  we marginalise over $A$, and the last equality holding since $ Y(0)\ci Z\vert W$ (Assumption ii').

Model $\mM$ thus implies
\begin{equation}\label{2ndstage}
E[Y\vert Z,W] = \omega_0(W) + m_0(W) E[A\vert Z,W],
\end{equation}
where  $\omega_0(W)=E[Y - Am_0(W) | W]$ being equal to $E[\varpi_0(W,U) \vert W]$ or  $E[Y(0)\vert W]$, depending on whether model \eqref{struct_modY} or \eqref{IV-g1} is assumed.

Equation \eqref{2ndstage} implies
$E[Y\vert Z=1,W] - E[Y\vert Z=0,W] =  m_0(W) \left( E[A\vert Z=1,W] - E[A\vert Z=0,W]\right),$ therefore for a binary IV, under $\mM$ and the (conditional) IV assumptions, $m_0(W)$ is identified by
\begin{equation}\label{m_id}
m_0(W)=\frac{E[Y\mid Z=1,W]-E[Y\mid Z=0,W]}{E[A\mid Z=1,W]-E[A\mid Z=0,W]},
\end{equation}

Estimation of the conditional expectations in equation \eqref{m_id} would typically involve specifying  models for the mean exposure $E[A\vert Z,W]$ and the mean outcome $E[Y\vert Z,W]$. 

Denote by  $\omega(W)$ the model for $E[Y - Am_0(W) | W] $ and  $\pi(Z,W)$ the model for  $E[A\vert Z,W]$. Finally,  let $\mu(Z,W)$ denote the implied model for $E[Y\vert Z,W]$.

\section{Doubly robust estimation for the linear instrumental variable model}\label{Sec:DR}

 To illustrate the methods, we consider throughout a situation where interest lies in the main effect modification by a single variable  $V\in W$, with a working parametric model for the treatment effect curve as a function of this single variable being:
\begin{equation}
\label{Parametric_m0}
m(W;\psi)=\psi_c + \psi_v V.
\end{equation}
The statistical parameter of interest  is therefore $\psi=(\psi_c, \psi_v)$, where $\psi_c$ represents the main causal treatment effect, and $ \psi_v$ is the effect modification by $V$. The function $m(W;\psi)$ can be interpreted as a working model for $E[m_0(W)\vert A=1, V]$. Importantly, the working parametric model  is not assumed to be the true model for $m_0(W)$.

 \subsection{TSLS}\label{Sec:TSLS}
Estimation of the expectations in equation \eqref{m_id} is often done  via an approach known as two-stage least squares (TSLS). The first stage fits a linear treatment selection model, that is a model for $A$ conditional on the instrument and the baseline covariates of interest, and then, the second stage is a linear model for eq. \eqref{2ndstage}, that is a linear regression for the outcome on the predicted treatment received and baseline covariates. We write
\begin{eqnarray}\label{TSLS}
\label{1TSLS} E[A\vert Z,W] &= &\pi(Z,W),\\
\label{2TSLS}E[Y\vert Z,W] &=& \omega(W) + m(W) \pi(Z,W).
\end{eqnarray}
In principle, there are many parametric choices for the second stage models, $\omega(W)$ and $m(W)$.  For TSLS to be consistent, the first stage model   $\pi(Z,W)$ must  be the parametric linear regression implied by the second stage, i.e. it must include all  the covariates and interactions appearing in the second stage model.

For example, if we assume working models $m(W;\psi)=\psi_c + \psi_v V$, and  $\omega(W;\beta)= \beta^\top W$, where abusing notation we assume the vector of ones is the first column of $W$, then the first-stage would  involve two equations, as follows 
\begin{eqnarray}\label{TSLS-EF}
\nonumber E[A\vert Z,W] &= &  \alpha_z Z +\alpha_{zv} ZV  +\alpha_{v} V  +   \sum_{W_i\in W\setminus V} \alpha_{wi} W_i,  \\
E[AV\vert Z,W] &=&  \lambda_ z Z + \lambda_{zv} ZV +  \lambda_v V   + \sum_{W_i\in  W\setminus V} \lambda_{wi} W_i,
\end{eqnarray}
where again,  $W$ includes 1 to allow for an intercept.
Because estimation of these two first-stage models is done separately without acknowledging that the model for $A$  should imply the model for $AZ$, the resulting TSLS estimator may be inefficient \citep{Vansteelandt2015}.  

 \citet{Vansteelandt2015} show that  standard TSLS estimation of $\beta$ and $\psi$ in equation  \eqref{2TSLS} is equivalent to solving an estimating equation of the form
\begin{equation}\label{ee2TSLS}
0=\sum_{i=1}^n 
 e_y(Z_i,W_i) \left\{Y_i-\omega(W_i;\beta)-m(W_i;\psi) \pi(Z_i,W_i;\alpha_0)\right\},
\end{equation}
for a given $\alpha_0$, where $ e_y(Z_i,W_i)$ is any conformable index vector function  of dimension $\mbox{dim}(\beta) + d$.

The estimators $\hat{\beta}$ and $\hat{\psi}$ obtained solving equation \eqref{ee2TSLS}, after substituting $\hat{\alpha}$ for $\alpha_0$ (the estimator from the first stage),  are consistent asymptotically normal (CAN), when both models $\omega(W;\beta)$ and $m(W;\psi)$ are correctly specified, i.e. even when $\pi(W;\alpha)$, the first stage model for the exposure, is misspecified \citep{Robins2000,Wooldridge2010}. Moreover, in the specific settings where the treatment effect curve $m(W;\psi)$ is linear in the covariates and the instrument is independent of $W$, TSLS is also robust to misspecification of $\omega(W;\beta)$. We refer the interested reader to  \citet{Vansteelandt2015}, Appendix B Proposition 5, for the proof.

This means that for estimators which are doubly robust in the more general settings, with either a treatment effect model  $m(W;\psi)$ that depends on the covariates (treatment effect heterogeneity),  or where the instrument $Z$ depends on $W$, methods beyond TSLS need to be considered.

\subsection{G-estimation}\label{Sec:gest}

Thus far, we have shown that in treatment effect modification settings with a binary conditional IV, the TSLS  IV estimator is consistent if the treatment-free outcome model $\omega(W)$ is correctly specified. An approach to obtaining  a doubly robust estimator involves modelling $E[Z\vert W_i]$ in structural nested mean models \citep{Robins1994}. Then,  the parameter of interest $\psi$ can be estimated using G-estimation. 

G-estimation exploits the idea that,  on average, there is no residual association between $Z$ and $E[Y - Am_0(W) | W]$. This suggest an estimation strategy for finding the parameters that make the empirical conditional covariance between $Z$ and the treatment-free potential outcome $Y(0)$ equal to 0. 
The resulting estimator is consistent if either the model for the conditional expectation $E(Z|W)$ or the treatment-free outcome model $\omega(W)$ or both are  correctly specified, and the assumption that partially linear IV model $\mM$ for the conditional mean of Y given $W$ and $Z$ is correct.
  The model for the conditional distribution of the binary IV  $g_0(W;\gamma_0)=E[Z\vert W]= P_0(Z=1\vert W)$ is often called  the instrument propensity score, and it is assumed to be a known function of $W$, smooth in a finite dimensional parameter $\gamma_0$.  

\citet{Okui2012} showed that this g-estimator for $\psi=(\psi_c, \psi_v)$ can be obtained 
as a solution to the following estimating equation \citep{Okui2012}
\begin{equation}\label{IV-geq}
0=\sum_n 
\left(e(Z_i,W_i; \gamma_0)-E[e(Z_i,W_i;\gamma_0)\vert W_i]\right)\left\{Y_i-\omega_0(W_i;\beta_0)-m_0(W_i;\psi)A_i\right\},
\end{equation}
where $e(Z,W)$ is any conformable vector function, i.e. of the appropriate dimension $\mbox{dim}(\beta) + d$, with $d=\mbox{dim}(\psi)$.

This  can be made (locally) efficient by choosing \citep{Vansteelandt2015}
\begin{equation}\label{conf_funct}
e(Z,W;\gamma_0) = \sigma_0^{-2}(Z,W)\left(\begin{array}{c}
1\\ V\end{array}\right)\left[ \mao(Z,W; \alpha_0)- \frac{E[\sigma_0^{-2}(Z,W)\mao(Z,W; \alpha_0)\vert W]}{E[\sigma_0^{-2}(Z,W)\vert W]} \right]
\end{equation}
  and
 $\sigma_0^{2}(Z,W)=\mbox{Var}\{Y-Am(W;\psi)\vert Z,W\}$. 

This estimator requires the user to specify working parametric models for $E(Y-Am(W;\psi)\vert W)$ and $E(Z\vert W)$,  i.e. to specify working models for $\omega(W;\beta)$ and $g(W;\gamma)$. The estimator  (denoted by IV-g) considered here  estimates both the parameter of interest $\psi$ and the nuisance parameter $\beta$ jointly, though approaches that estimate $\beta$ consistently first have also been proposed \citep{Okui2012}. This  can be made feasible by replacing $\alpha_0$, $\beta_0$ and $\gamma_0$ by their corresponding consistent estimators $\hat{\alpha}$, $\hat{\beta}$ and $\hat{\gamma}$, and setting $\sigma_0^2$ equal to 1 (as it is just a proportionality constant).  It has been shown to be CAN if either the model for $E(Y-Am(W;\psi)\vert A, W)$ or the model for $E(Z\vert W)$ is correct, and hence consistent whenever the model for $m_0(W)$ is correctly specified \citep{Okui2012}. The addition of the instrument propensity score model to the TSLS estimating equations \eqref{ee2TSLS} is particularly helpful when the  dependence between $Z$ and the baseline covariates is known, as would be the case in a randomised trial,  thus guaranteeing robustness against misspecification of the outcome model.

Moreover, the IV-g estimator is efficient when  all three models are correctly specified \citep{Vansteelandt2015}.

The influence function of the IV-g estimator can be written as
\begin{equation}
D_i(\psi)(O_i)= M(Z_i,W_i,A_i)^{-1}K(Z_i,W_i) \left(\begin{array}{c}
1\\ V_{i}\end{array}\right) \{ Y_i - \omega(W_i;\beta)\} - \left(\begin{array}{c}
\psi_{c}\\ \psi_{v} \end{array}\right) 
\end{equation}
with
\begin{equation}
 K(Z,W)=\pi(Z,W; \alpha)- E_{g( W;\gamma_0)}[\pi(Z, W;\alpha)\vert W] 
\end{equation}
and 
\begin{equation}
M(Z,W,A) = A K(Z,W) \left(
\begin{array}{cc}
1& V\\ V & V^2
\end{array}\right).
\end{equation}

Since the IV g-estimator is CAN, the asymptotic variance is the variance of its influence function, i.e. ${\mbox{Var}}(\psi)=E[D(\psi)^\top D(\psi)]$ \citep{Newey1990}. Therefore, we obtain an estimate of the variance by the sample variance of the estimated influence function, obtained by plugging-in  consistent estimators for $\alpha$, ${\beta}$ and ${\gamma}$, 
$$\widehat{\mbox{V}}(\widehat{\psi})=n^{-1}\mbox{Var}_n( \widehat D(\widehat\psi)),$$ where we have used the subscript $n$ to denote the sample variance on a sample of size $n$.  
This variance estimator  ignores the nuisance parameter estimation. Robust standard errors can be obtained via  the bootstrap or a sandwich estimator.

The IV-g estimator  gains efficiency from assuming that the working model for  the treatment effect curve, equation \eqref{Parametric_m0}, $m_\psi(W)=m(W;\psi)=\psi_c + \psi_v V $, holds when this is correct. 
However when model  \eqref{Parametric_m0}  is misspecified (e.g. that the treatment effect curve depends on more covariates, not just $V$, or that the relationship is not linear), the IV-g estimator will behave as a projection onto the working parametric model, so long as the mean exposure model 
$\mao(Z,W)$ is correctly specified  and  
$\mbox{Cov}(\left\{\mao(Z,W)-E(\mao(Z,W)\vert W)\right\},A\vert W)$ is constant in $W$.  

\subsection{Data-adaptive  estimation}\label{Sec:dataadapt}

The IV-g estimator presented thus far is restricted to using parametric working models for the nuisance parameters. Since  all working models are  likely to be misspecified in practice, the resulting estimator is unlikely to be consistent.  

An increasingly popular strategy to avoid the bias introduced by such model misspecification and have valid inferences is to use machine learning estimators for the nuisance parameters. 
This is made possible since DR estimators can converge at fast rates ($\sqrt{n}$) to the true parameter, and are therefore CAN, even when the nuisance functionals have been  estimated via machine learning, under either empirical process conditions (e.g. Donsker class)  restricting the complexity of the nuisance functionals,  or using sample splitting \citep{van2011targetedbook, Farrell2015, Chernozhukov2017AER, Kennedy2017, Athey2018}.

Briefly, if the score function $\mathcal{S}$ of the DR estimator is \textit{Neyman orthogonal} to the nuisance parameters i.e. the path-wise (or Gateaux) derivative of the score function exists and vanishes at the true value of the nuisance parameters, then, as long as the data-adaptive estimators for all nuisance functionals converge to their respective truths, and  the product of their convergence rates is faster than ${n}^{-\frac{1}{2}}$,  the DR estimator is CAN and inference can be based on the IF. Convergence rates for these data-adaptive estimators depend on the smoothness and number of covariates included \citep{Gyoerfi2006}.

Machine learning estimation of the nuisance parameters of DR estimators for the partially linear IV model has been studied recently.   \citet{Chernozhukov2016} give sufficient conditions to guarantee that using data-adaptive fits for the nuisance functionals in DR estimators constructed from estimating equations based on Neyman-orthogonal scores results in valid inferences. In particular,  consider the score function 
\begin{equation}\label{scoreC1}
\mathcal{S}_i= \left\{Z_i-g(W_i)\right\}\left\{Y_i-\omega(W_i)-m(W_i;\psi)A_i\right\},\end{equation}
where $g(W_i)$ and $\omega(W_i)$ are  $L^2$-functions  with respect to $P_0$, mapping $W\mapsto \mathbb{R}$.  Assuming $Y, A$ and $Z$ are bounded and with finite variance bounded away from zero,  the estimator obtained as a solution to the estimating equation with score \eqref{scoreC1}
 is CAN even after plugging in data-adaptive nuisance estimators, as long as these satisfy: 
 \begin{equation}\label{ratesIV}
\norm{ \hat{g}(W)- g_0(W)} \times \norm{ \widehat{\omega}(W)- \omega_{0}(W)}
<o_{P}({n}^{-\frac{1}{2}}),
\end{equation}
where $\norm{ \circ}=\norm{ \circ}_{P,2}$ i.e. the $L^2$-functions  with respect to $P_0$.

  We refer the interested reader to \citet{Chernozhukov2016} for the technical details. 

Since the g-estimator for the IV model is Neyman orthogonal, data-adaptive IV-g estimators can be obtained by solving equation \eqref{IV-geq} after data-adaptive estimates for $\omega(W)$ (the treatment-free outcome model), $\pi(Z,W)$ (the exposure model) and/or  $g(W)$ (the instrument propensity score) have been plugged in.  Under sufficient regularity conditions, and provided the data-adaptive models used converge sufficiently fast to their respective true parameter, the resulting IV g-estimator is CAN.

For example, solving equation \eqref{IV-geq} where we have plugged in fits from a parametric model for $\pi(Z,W)$ and data-adaptive estimates for $E[Y - Am_0(W) | W]$  and $E[Z\vert W]$,  the arguments used in  \citet{Chernozhukov2016} can be applied directly to show that  the IV-g estimator is CAN when eq. \eqref{ratesIV} holds.  To see why, consider a parametric model for $\pi(Z,W)=\alpha_0 + \alpha_1 Z + \alpha_2 V$, so that the score function for eq. \eqref{IV-geq} is $\mathcal{S}=(1\quad V)^\top \left\{ \alpha_1 \left(Z-g(W)\right)\left( Y-\omega(W)-Am(W;\psi) \right)\right\}$, which is of the form eq. \eqref{scoreC1}.

The data-adaptive IV-g estimator implemented here uses data-adaptive estimates for  $E[A\vert Z,W$ and $E[Z\vert W]$  but estimates jointly the parametric $m(W;\psi)$ and $\omega(W,\beta)$ as before. 
The resulting estimator can be shown to be CAN if the nuisance models converge to their respective truths at the rates of convergence in equation \eqref{ratesIV},  under sufficient regularity conditions. See the Appendix for a sketch of the proof. 

To obtain  the data-adaptive estimates, we use the Super Learner (SL)  \citep{van2007super}.  The SL uses cross validation to  find the optimal weighted convex combination of  multiple candidate estimators specified by the user in the SL \textit{library}. The library can include parametric and non-parametric estimators. The SL has been shown to perform asymptotically as well as the best learner included in its library, so that  adding additional algorithms improves the performance of the SL.  The finite sample performance of the SL has been demonstrated extensively in simulations \citep{van2007super, Porter2011, Pirracchio2015}. 
The use of data-adaptive fits for nuisance functionals has been extensively exploited within the TMLE literature which  we review next. 

\subsection{Targeted minimum loss estimation (TMLE)}\label{Sec:TMLE}

Targeted minimum loss estimation (TMLE) is a general approach for causal inference, which has been adopted on a wide range of causal problem 
\citep{gruber2010targeted,van2011targetedbook,zheng2012targeted,van2012targeted,petersen2014targeted}. 

TMLE is a semi-parametric influence-function based estimation approach, which incorporates a ``targeting'' step that guarantees the resulting estimator has a well behaved higher-order residual term.  Most commonly, it  combines estimates of nuisance functionals and an initial estimate of the target parameter. 
 These initial estimates can be obtained by specifying parametric models or, under empirical processes conditions (e.g. Donsker class) which can be relaxed using sample splitting \citep{Zheng2011},  via machine learning. Typically, the TMLE literature uses the Super Learner with cross-validation \citep{van2007super}.  Assuming the  data-adaptive estimates converge to their respective truths sufficiently fast, the resulting TMLE is CAN. We refer the interested reader to \citet{van2011targetedbook} and \citet{van2018targetedbook}. 

\citet{Toth2016} proposed three TMLE  estimators  for the  (partially) linear IV model.  In the next section, we describe in more detail the non-iterative linear TMLE, which we denote  by IV-TMLE.

\subsubsection{IV-TMLE}\label{Sec:tmle}

Let $\Psi:\mP\mapsto \R^2$ be the target parameter mapping from the space of all possible models for the true distribution of the data $P_0$  to $\R^2$, defined by projecting the treatment effect curve onto the working parametric model $m_\psi=\psi_c + \psi_v V$, i.e.
  $\Psi(P_0)=(\psi_c, \psi_v)=\psi_0$  is the solution to $$E\left[ \left(\!\!\begin{array}{c}1\\ V\end{array}\!\!\right)\{m_0(W)-(\psi_c + \psi_v V)\}\,\right]=0.$$
We note that $\Psi$ only depends on $P_0$ through $m_0$ and the distribution of $Z$ and the covariates $P_{0W}$. We denote this relevant part by $Q_0=(m_0,g_0, Q_{0W})$ with $Q_{0W}=P_{0W}$.

Under the IV model $\mM:\ E[Y\vert Z,W] = \omega_0(W) + m_0(W) \mao(Z,W)$, the treatment effect curve $m_0(W)$ depends on $\mu_0(Z,W)=E[Y\vert W,Z]$ and $\mao(Z,W)$, and thus construction of a TMLE for the IV model starts by  obtaining initial estimates of  $\mu(Z,W)$, $\pi(Z,W)$, and the instrument propensity score  $g(W)$. We denote these initial estimates by a $0$ superscript. From these, and model \eqref{2ndstage}, we calculate an initial estimate  for $m(W)$, denoted $m^0(W)$.

The next step in the construction of a TMLE requires the specification of a loss function $L(P)$, such that the expectation of the loss function is minimised at the true probability distribution, $E_{0}[L(P_0)(O)]=\mbox{min}_{P\in \mP} E_{0}(L(P)(O))$. Here,  we use the square error loss function. 
Under the IV model $\mM$ and the working model for the treatment effect  curve $m_\psi(V)=\psi_c + \psi_v V $, the efficient influence function (EIF) can be written as:
\begin{eqnarray}\label{EIF-TMLE}
\nonumber
&&D^*(m,g,Q_W)(O)=h(W)\{\mao(Z,W) - E_{0}(\mao(Z,W)\vert W)\} \{Y-\mao(Z,W)m_0(W) -\omega_0(W)\} \\
&&\qquad\quad-
 h(W)\{\left(\mao(Z,W)\!-\!E_{0}[\mao(Z,W)\vert W]\right) m_0(W)\}\left(A-\mao(Z,W) \right) + D_W(Q_{W}),
 \end{eqnarray} 
where   $h(W)$ is the so-called \textit{clever covariate}, defined as
\begin{equation}\label{cleverh}
 h(W)= \mbox{Var}(V)^{-1}\left(\!\!\!\!\begin{array}{c}
 E[V^2]-E[V]V\\  V-E[V]\end{array}\!\!\!\right) \zeta^{-2}(W)
\end{equation}
with the term $\zeta^2(W)$, which is associated with instrument strength, being
\begin{eqnarray}\label{eq_zeta}
\nonumber\zeta^2(W)&=&  \mbox{Var}_{Z\vert W}\left(\pi(Z,W)\vert W\right),\\
\nonumber&=& E[\{ \pi(Z,W) - \sum_{z\in \{0,1\}} \pi(z,W)g(Z=z,W)\}^2 \vert W],\\
\nonumber&=& \left\{\pi(1,W)-\pi(0,W)\right\}^2g(W)(1-g(W)),\\
&=& \left\{\pi(1,W)-\pi(0,W)\right\}^2\mbox{Var}(Z\vert W).
\end{eqnarray}

Finally $D_W(Q_{W})=c \,\left\{ m_0(W) -m_\psi(V) \right\}$.

The \textit{targeting} step involves fitting  a linear model for $m(W)$ on the single ``clever'' covariate $h(W)$ with the initial  estimate $m^0(W)$ as an offset,
\begin{equation}\label{lin_fluct}
m^*(\epsilon)(W)={m}^0(W) + h(W)^T \epsilon.
\end{equation}

Estimation of the coefficient in equation \eqref{lin_fluct} involves solving the empirical EIF equation,
\begin{equation}\label{eps}
\frac{1}{n}\sum_{i=1}^n D^*(m^*(\epsilon),g^0, Q_W)(O_i)=0,
\end{equation}
or equivalently, solving for $\epsilon$ a system of $d$ linear equations:
\begin{equation*}
\frac{1}{n}\sum_{i=1}^n  h^0(W_i)\{  \pi^0(Z_i,W_i)- E_{g^0(W_i)}[\pi^0(Z_i,W_i)\vert W_i] \}\left\{Y_i- A_i\left(m^0(W_i)+h^0(W_i)^\top \epsilon\right)-\omega^0(W_i) \right\}=0,
\end{equation*}
where  $h^0(W_i)$ is obtained by  plugging in $\pi^0(Z_i,W_i)$ and $g^0(W_i)$ into the equations defining the clever covariate \eqref{eq_zeta} and \eqref{cleverh}.

Denote by $\epsilon^*$  the solution to  equation \eqref{eps}. Then, the non-iterative linear TMLE estimator of $m_0(W)$ is obtained by substituting $\epsilon^*$ into equation \eqref{lin_fluct}. Finally, we  project  the resulting function $m^*(\epsilon^*)(W)$ onto the working model $m_\psi$ by OLS,  obtaining $(\psi_c^*,\psi_v^*)$, the TMLE estimator of the statistical parameters of interest.

 \citet{Toth2016} showed that this approach results in an estimator which is double-robust, i.e. consistent  when (i) the initial estimators of $\pi_0(Z,W)$ and $g_0$ are consistent, (ii) the initial estimators of  $m_0$ and $g_0$ are consistent, or (iii)  the initial estimators of  $m_0$ and $\omega_0$ are consistent. However, using a linear fluctuation model has the drawback that the resulting estimates are not guaranteed to be constrained within the bounds implied by the data.

We remark that the variance of the IV-TMLE estimators becomes very large when the term $\zeta^2(W)$ is very small. Since $\zeta^2(W)$ captures the strength of the
instrument in predicting the exposure given $W$, the IV-TMLE estimators become unstable with large variance when the instrument is weak. To stabilise the estimators, we choose the maximum of the estimated value of $\zeta^2(W)$ and 0.025 when constructing the clever covariate for a given data set.

\section{Simulation Study}\label{Sec:Simulation} 

 We perform a factorial simulation study to assess the finite sample performance of the alternative methods to estimate the statistical parameter of interest, under the different combinations of $\omega$, $\pi$ or $m$ being in turn correctly specified or not, while the instrument model is always correct. We write $\mathbb{1}(k \neq {k_0})$ as an indicator function for scenarios where the assumed model for $k\in\{\omega,\pi,m\}$  is misspecified.

We  generate  data to mimic a randomised controlled trial with two-sided non-adherence, i.e. both randomly allocated groups have a non-zero probability of receiving the opposite treatment. The are two different sample sizes, small $n=500$ and large $n=10,000$.
We begin by generating five independent standard normal variables $W_1, \ldots, W_4$ and $V$. These are the observed baseline covariates, of which one is the effect modifier $V$. We also generate a standard normal unobserved confounder $U$. 
We generate randomised treatment  also independently of the other variables, $Z \sim \mbox{Bern}(0.6)$, and then simulate the binary treatment received  $A \sim \mbox{Bern}(\pi_0(W,V,U,Z))$, i.e. the probability of getting the active treatment is a function  the baseline variables, the unobserved confounder, and the instrument,  namely
$$\mbox{logit}(\pi_0)=1.5 Z + 0.03 V + 0.01 W_1 +0.01 W_2 + 0.01 W_3+ 0.01 W_4  + 0.03U - \mathbb{1}(\pi \neq {\pi_0}) (5 ZW_1 ).$$

Notice that we are generating the exposure $A$ in such a way that the condition necessary for the IV-g estimator to converge to the parameter of interest when $\mM$ is wrong is no longer satisfied, for settings where the true $\pi_0\neq\pi$.

 The  continuous outcome $Y$ is then simulated  from $Y \sim N(\mu_0,1 )$, with $\mu_0$ given by:
  \begin{eqnarray*}
\mu_0&=&\omega_0(W)  + m_0(W) A  + U,\\
\omega_0(W)&=&\{1-\mathbb{1}(\omega \neq \omega_{0})\}\{0.5 + 0.5V + 0.01W_1 +0.01W_2 + 0.01W_3+ 0.01 W_4 \}  \\
      &+&         \mathbb{1}(\omega \neq \omega_{0})\mbox{exp}\{0.05 + 0.05 V + 0.001W_1 + 0.001W_2 + 0.001 W_3 \\
&&\qquad\qquad +\ 0.001W_4-  \ 0.2V(W_1 +W_2 + W_3 + W_4)\},\\
m_0(W)&=&0.5 + 0.5  V + \mathbb{1}(m \neq m_{0})\{3(W_1 + W_2 + W_3+ W_4)\},
      \end{eqnarray*}
which means that the true $E[m_0(W)\vert  A=1, V=v]=0.5 + 0.5 V$, i.e. $\psi_0=(\psi_{c0},\psi_{v0})=(0.5,0.5)$.

We generate 1,000 replicates for each scenario.  We perform analyses with TSLS,  IV-g and IV-TMLE, the latter two are implemented with and without the data-adaptive estimation of  nuisance models. 
For  the TSLS, the first stage is  as per equation \eqref{TSLS-EF}.  Parametric IV-g and TMLE use main terms logistic models for the instrument propensity score and the treatment model, namely 
$$\mbox{logit}(g(W;\gamma))=\mbox{logit}\{P(Z=1\vert W)\} =  \gamma_0 +  \sum_{i=1}^4 \gamma_{i} W_i+ \gamma_5V,$$ and 
$$\mbox{logit}(\pi(W,Z;\alpha))=\mbox{logit}\{P(A=1\vert Z,W)\}=  \alpha_z Z  +   \sum_{i=1}^4 \alpha_{i} W_i + \alpha_5V$$.

For the data-adaptive estimation of $\pi(Z,W)$ and $g(W)$, we use the Super Learner. Since $A$ and $Z$ are binary, the library used includes glm (generalised linear models), step (stepwise model selection using AIC),   svm (support vector machines, with radial basis functions) and gam (generalised additive models), with  linear and second-order terms used for the glm, step and gam learners. 

In addition, for  the IV-TMLE, we use need data-adaptive estimates of the continuous outcome. The  library of learners used for $\mu(Z,W)$ and $\omega(W)$ includes  glm, step, svm and polymars (multivariate adaptive polynomial spline regression), chosen in order to preserve the linear structure of the partially linear IV model \eqref{2ndstage}.

The SEs of the parametric  IV-g and TMLE  are obtained by bootstrapping (percentile 95\% confidence intervals (CI) using 1999 bootstrap samples), while for the data-adaptive estimators the SEs are based on the empirical variance of the estimated (E)IF. 

We compute the mean bias of the estimates, coverage of  95\% CI,  and root mean square error (RMSE).  

\subsection{Results from the simulation}

\begin{figure}
\caption{Performance (Bias and  Coverage) of TSLS, TMLE and IV-g estimators, when the sample size is $n=500$. Scenarios with correct or misspecified $\pi$ and $\omega$ vary by column,  $m$ correctly specified is plotted in black while  $m$ misspecified is plotted in grey. The hollow shapes  correspond to parametric nuisance models estimation, and the solid shapes correspond to estimators using data-adaptive nuisance model estimates. The bias is presented with Monte Carlo Error CIs. Results corresponding to bias $\geq 2$ in absolute value are not plotted, but can be found in Table \ref{tab:excluded}. Dotted line in the bias plot is the 0 line, the dashed lines  in the coverage plot are the 92.5 and 97.5 \% coverage rates.}
\label{fig:bias500}
\begin{center}
\includegraphics[scale=.9]{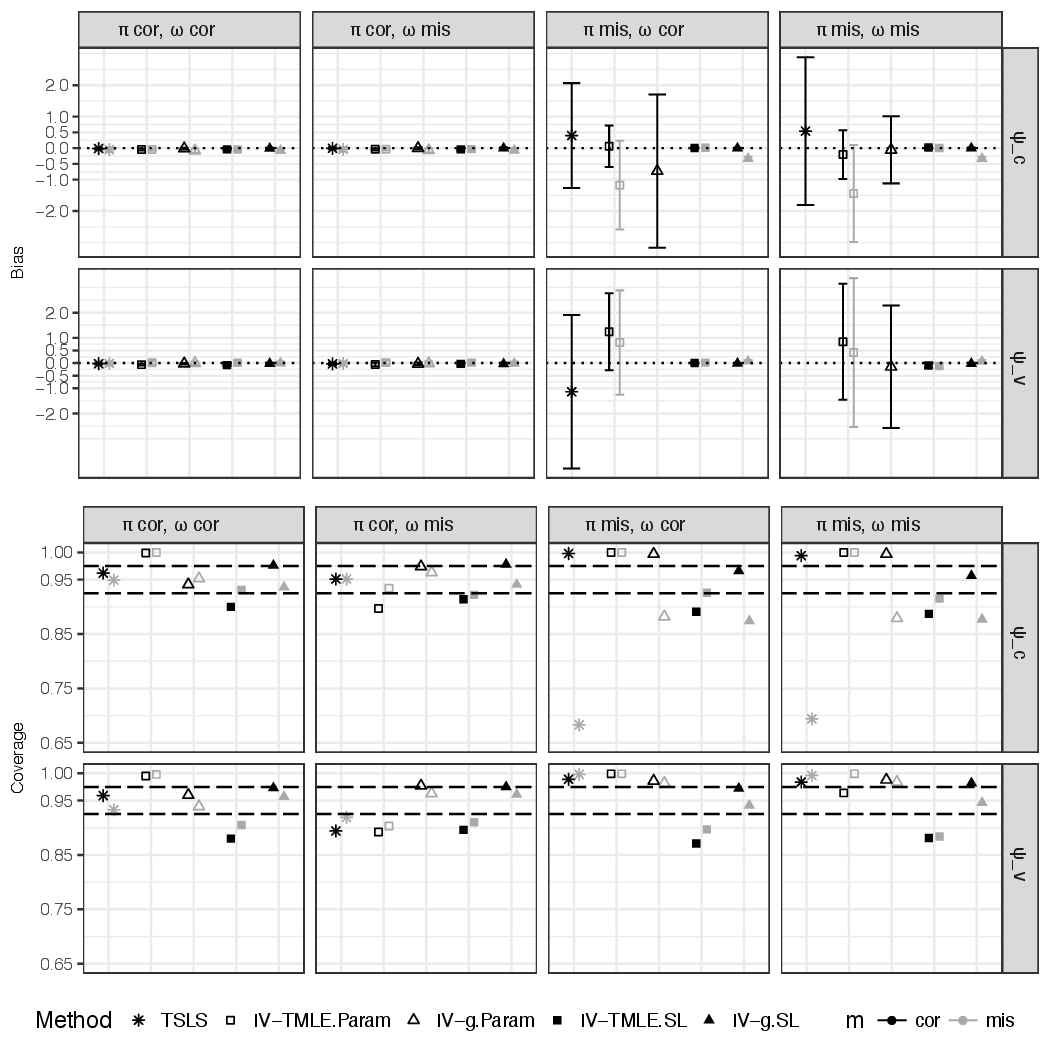}
\end{center}
\end{figure}

\begin{figure}
\caption{Performance (Bias and  Coverage) of TSLS, TMLE and IV-g estimators, when the sample size is $n=10,000$. Scenarios with correct or misspecified $\pi$ and $\omega$ vary by column,  $m$ correctly specified is plotted in black while  $m$ misspecified is plotted in grey. The hollow shapes  correspond to parametric nuisance models estimation, and the solid shapes correspond to estimators using data-adaptive nuisance model estimates. The bias is presented with Monte Carlo Error CIs. Results corresponding to bias $\geq 2$ in absolute value are not plotted, but can be found in Table \ref{tab:excluded}. Dotted line in the bias plot is the 0 line, the dashed lines  in the coverage plot are the 92.5 and 97.5 \% coverage rates. rates.}\label{fig:bias10K}
\begin{center}
\includegraphics[scale=.9]{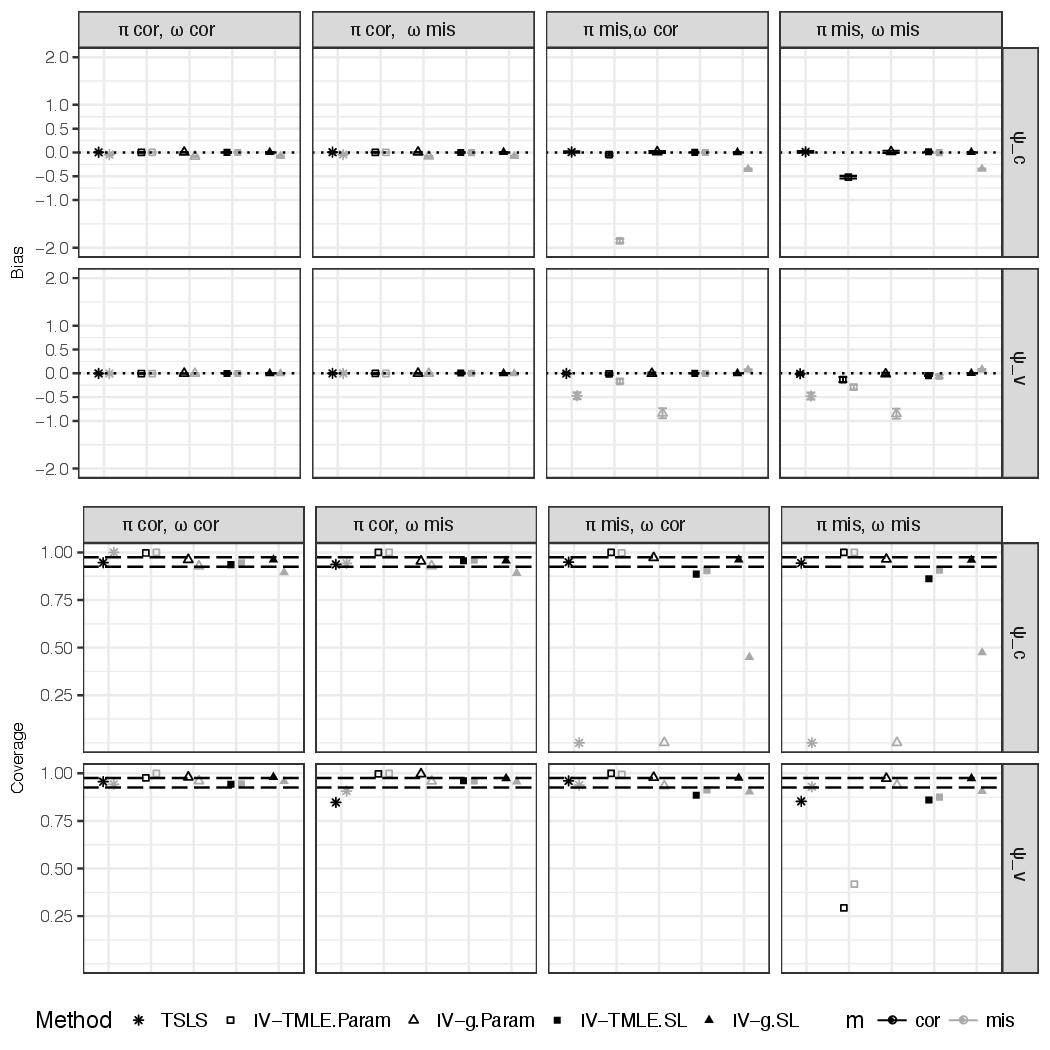}
\end{center}
\end{figure}

Figures \ref{fig:bias500}  and  \ref{fig:bias10K} show the mean bias  (top) and CI coverage rate  (bottom) corresponding to scenarios with  sample size of $n=500$ and  $n=10,000$ respectively. For clarity, the figures show only the methods resulting in absolute bias less than 2 are plotted, corresponding to those having absolute bias $<400\%$ of the true parameter value. The  excluded results are reported in Table \ref{tab:excluded} in the Appendix.

When all models are correctly specified (first column, plotted in black), all methods show close to zero bias for both of the target parameters. At large sample sizes ($n=10,000$), the coverage levels are close to the nominal value (between 92.5 and 97.5\%) for TSLS and IV-g estimator. In contrast, the bootstrapped CIs corresponding to parametric TMLE  result in over-coverage (99\%), while the EIF-based CIs for the data-adaptive TMLE shows under-coverage, which is especially in the small sample size scenarios ($n=500$), dropping below 90\% for $\psi_v$. This low-coverage phenomenon of the EIF-based CIs for TMLE estimators has been noted before by  \citet{vdLGruber2011} and \citet{petersen2014targeted}.

TSLS performs well when $m$ and $\pi$ are correctly specified (second column), but when the exposure model $\pi$ is misspecified (3rd and 4th columns), it performs poorly, even in scenarios where $m$ and the IV model are correctly specified (plotted in black), demonstrating numerically the lack of double robustness. When $m$ is misspecified (plotted in grey), TSLS results in bias $\geq 200\%$ of the true effect (not plotted in the Figures, see  Table \ref{tab:excluded}).  Consequently,  the coverage of the CIs is poor, being close to 0 in the larger sample size settings.
 
Both parametric TMLE and g-estimator result in small levels of bias and good coverage under those misspecified scenarios when the double robust properties are expected to provide protection. For example, with $m$ correctly specified, the g-estimator has small bias and good coverage even when the exposure model $\pi$ and the outcome model $\omega$ are misspecified (the last column of the Figures). TMLE on the other hand shows some significant bias, even at large samples $n=10,000$. However, implementing the IV-g and IV-TMLE methods using the Super Learner returns the bias  and coverage to the levels reported under correct specification, with TMLE still showing coverage under 92.5\%. 

Where $m$ is misspecified, we would expect the g-estimator to behave as a projection of the true treatment effect curve onto the working parametric model $m_\psi(W)$, as long as the model for the exposure $\pi(Z,W)$ is correctly specified and it is such that $\mbox{Cov}(\left\{\mao(Z,W)-E(\mao(Z,W)\vert W)\right\},A\vert W)$ is constant in $W$. 
Since the data generating models are such that the true $\pi_0(W,Z)$ has constant covariance with the received exposure given $W$,  we can see in the first two columns of Figures \ref{fig:bias500}  and  \ref{fig:bias10K},  that the g-estimator performance is adequate when the parametric model $\pi(W,Z)$ is correctly specified (empty triangles plotted in grey). In contrast, for scenarios where the true $\pi_0(W,Z)$ does not have conditional constant covariance with $A$ given $W$ (third and fourth columns), there is substantial remaining bias even after using data-adaptive fits for the nuisance models, especially for the intercept $\psi_c$ (see for example, in the last column of Figure \ref{fig:bias10K} plotted in grey). 

Tables \ref{tableRMSE500} and \ref{tableRMSE10K} report the RMSE results. When $m$ is correctly specified, IV-g outperforms all other methods, with the smallest RMSE. Where the working parametric model for the treatment effect curve $m(W)$ is misspecified, TMLE has smaller RMSE in most settings. Looking at the larger sample $n=10,000$, we can conclude that both DR estimators have reported performance according to their theoretical double-robust properties,  and the TSLS method showed similar performance to the parametric implementation of the IV-g method. Both DR methods have benefitted from the data-adaptive estimation of the nuisance parameters: the performance of the estimators have not been harmed in the correctly specified scenarios, and RMSE has been greatly reduced in the scenarios when the DR properties do not provide protection against misspecification.

\begin{table}
  \centering
  \caption{RMSE of the TSLS, TMLE and IV-g estimators, when the sample size is $n=500$.}\label{tableRMSE500}
    \begin{tabular}{rrrlcc}\hline\hline
    \multicolumn{1}{l}{Scenario} & \multicolumn{1}{l}{Nuisance models estimation} & \multicolumn{1}{l}{Parameter} & Method & \multicolumn{2}{c}{RMSE} \\  \hline
          &       &       &       & \multicolumn{1}{c}{$m(W)$ correct} & \multicolumn{1}{c}{$m(W)$ mis} \\ \hline
    \multicolumn{1}{l}{$\pi$ cor, $\omega$ cor} & \multicolumn{1}{l}{Parametric} & \multicolumn{1}{l}{$\psi_c$} & TSLS  & 0.446      & 1.030 \\
          &       &       & IV-g  & 0.443 & 1.084 \\
          &       &       & IV-TMLE & 0.473 & 0.606 \\ \cline{3-6}
          &       & \multicolumn{1}{l}{$\psi_v$} & TSLS  & 0.480 & 1.131 \\ 
          &       &       & IV-g  & 0.479 & 1.132 \\
          &       &       & IV-TMLE & 0.580 & 1.234 \\ \cline{2-6}
          & \multicolumn{1}{l}{SL} & \multicolumn{1}{l}{$\psi_c$} & IV-g  & 0.439 & 1.159 \\
          &       &       & IV-TMLE & 0.475 & 0.614 \\ \cline{3-6}
          &       & \multicolumn{1}{l}{$\psi_v$} & IV-g  & 0.468 & 1.117 \\
          &       &       & IV-TMLE & 0.586 & 1.160 \\
 \hline
    \multicolumn{1}{l}{$\pi$ cor, $\omega$ mis} & \multicolumn{1}{l}{Parametric} & \multicolumn{1}{l}{$\psi_c$} & TSLS  & 0.520 & 1.065 \\
          &       &       & IV-g  & 0.517 & 1.119 \\
          &       &       & IV-TMLE & 0.548 & 0.655 \\ \cline{3-6}
          &       & \multicolumn{1}{l}{$\psi_v$} & TSLS  & 0.782 & 1.314 \\
          &       &       & IV-g  & 0.788 & 1.338 \\
          &       &       & IV-TMLE & 1.073 & 1.262 \\ \cline{2-6}
          & \multicolumn{1}{l}{SL} & \multicolumn{1}{l}{$\psi_c$} & IV-g  & 0.495 & 1.183 \\
          &       &       & IV-TMLE & 0.616 & 0.685 \\ \cline{3-6}
          &       & \multicolumn{1}{l}{$\psi_v$} & IV-g  & 0.753 & 1.295 \\
          &       &       & IV-TMLE & 1.111 & 1.368 \\
       \hline \multicolumn{1}{l}{$\pi$ mis, $\omega$ cor} & \multicolumn{1}{l}{Parametric} & \multicolumn{1}{l}{$\psi_c$} & TSLS  & 26.835 & 160.523 \\
          &       &       & IV-g  & 39.241 & 446.003 \\
          &       &       & IV-TMLE & 10.649 & 22.791 \\ \cline{3-6}
          &       & \multicolumn{1}{l}{$\psi_v$} & TSLS  & 49.141 & 163.553 \\
          &       &       & IV-g  & 139.285 & 1596.750 \\
          &       &       & IV-TMLE & 24.685 & 33.392 \\ \cline{2-6}
          & \multicolumn{1}{l}{SL} & \multicolumn{1}{l}{$\psi_c$} & IV-g  & 0.316 & 0.990 \\
          &       &       & IV-TMLE & 0.472 & 0.557 \\ \cline{3-6}
          &       & \multicolumn{1}{l}{$\psi_v$} & IV-g  & 0.309 & 0.822 \\
          &       &       & IV-TMLE & 0.756 & 0.858 \\
\hline    \multicolumn{1}{l}{$\pi$ mis, $\omega$ mis} & \multicolumn{1}{l}{Parametric} & \multicolumn{1}{l}{$\psi_c$} & TSLS  & 37.825 & 154.812 \\
          &       &       & IV-g  & 17.157 & 420.882 \\
          &       &       & IV-TMLE & 12.496 & 24.872 \\ \cline{3-6}
          &       & \multicolumn{1}{l}{$\psi_v$} & TSLS  & 75.209 & 150.203 \\
          &       &       & IV-g  & 39.172 & 1491.963 \\
          &       &       & IV-TMLE & 37.065 & 47.649 \\ \cline{2-6}
          & \multicolumn{1}{l}{SL} & \multicolumn{1}{l}{$\psi_c$} & IV-g  & 0.367 & 1.011 \\
          &       &       & IV-TMLE & 0.743 & 0.788 \\ \cline{3-6}
          &       & \multicolumn{1}{l}{$\psi_v$} & IV-g  & 0.557 & 0.925 \\
          &       &       & IV-TMLE & 1.446 & 1.437 \\ \hline\hline
    \end{tabular}%
\end{table}%

\begin{table}
  \centering
\caption{ RMSE of TSLS, TMLE and IV-g estimators, when the sample size is $n=10,000$.}\label{tableRMSE10K}
 \begin{tabular}{rrrlcc}\hline\hline
      \multicolumn{1}{l}{Scenario} & \multicolumn{1}{l}{Nuisance models estimation} & \multicolumn{1}{l}{Parameter} & Method & \multicolumn{2}{c}{RMSE} \\  \hline
          &       &       &       & \multicolumn{1}{l}{$m(W)$ correct} & \multicolumn{1}{l}{$m(W)$ mis} \\ \hline
  \multicolumn{1}{l}{$\pi$ cor, $\omega$ cor} & \multicolumn{1}{l}{Parametric} & \multicolumn{1}{l}{$\psi_c$} & TSLS  & 0.092 & 0.207 \\
          &       &       & IV-g  & 0.092 & 0.228 \\
          &       &       & IV-TMLE & 0.092 & 0.112 \\ \cline{3-6}
          &       & \multicolumn{1}{l}{$\psi_v$} & TSLS  & 0.090 & 0.213 \\
          &       &       & IV-g  & 0.090 & 0.214 \\
          &       &       & IV-TMLE & 0.091 & 0.113 \\ \cline{2-6}
          & \multicolumn{1}{l}{SL} & \multicolumn{1}{l}{$\psi_c$} & IV-g  & 0.092 & 0.255 \\
          &       &       & IV-TMLE & 0.093 & 0.112 \\ \cline{3-6}
          &       & \multicolumn{1}{l}{$\psi_v$} & IV-g  & 0.090 & 0.215 \\
          &       &       & IV-TMLE & 0.093 & 0.114 \\ \hline
    \multicolumn{1}{l}{$\pi$ cor, $\omega$ mis} & \multicolumn{1}{l}{Parametric} & \multicolumn{1}{l}{$\psi_c$} & TSLS  & 0.107 & 0.213 \\
          &       &       & IV-g  & 0.107 & 0.234 \\
          &       &       & IV-TMLE & 0.107 & 0.125 \\ \cline{3-6}
          &       & \multicolumn{1}{l}{$\psi_v$} & TSLS  & 0.140 & 0.240 \\
          &       &       & IV-g  & 0.134 & 0.236 \\
          &       &       & IV-TMLE & 0.140 & 0.156 \\ \cline{2-6}
          & \multicolumn{1}{l}{SL} & \multicolumn{1}{l}{$\psi_c$} & IV-g  & 0.104 & 0.260 \\
          &       &       & IV-TMLE & 0.136 & 0.143 \\ \cline{3-6}
          &       & \multicolumn{1}{l}{$\psi_v$} & IV-g  & 0.133 & 0.237 \\
          &       &       & IV-TMLE & 0.206 & 0.163 \\ \hline
    \multicolumn{1}{l}{$\pi$ mis, $\omega$ cor} & \multicolumn{1}{l}{Parametric} & \multicolumn{1}{l}{$\psi_c$} & TSLS  & 0.270 & 10.160 \\
          &       &       & IV-g  & 0.333 & 9.660 \\
          &       &       & IV-TMLE & 0.349 & 1.957 \\ \cline{3-6}
          &       & \multicolumn{1}{l}{$\psi_v$} & TSLS  & 0.269 & 1.250 \\
          &       &       & IV-g  & 0.345 & 1.890 \\
          &       &       & IV-TMLE & 0.382 & 0.682 \\ \cline{2-6}
          & \multicolumn{1}{l}{SL} & \multicolumn{1}{l}{$\psi_c$} & IV-g  & 0.069 & 0.408 \\
          &       &       & IV-TMLE & 0.075 & 0.098 \\ \cline{3-6}
          &       & \multicolumn{1}{l}{$\psi_v$} & IV-g  & 0.066 & 0.196 \\
          &       &       & IV-TMLE & 0.076 & 0.100 \\ \hline
    \multicolumn{1}{l}{$\pi$ mis, $\omega$ mis} & \multicolumn{1}{l}{Parametric} & \multicolumn{1}{l}{$\psi_c$} & TSLS  & 0.317 & 10.158 \\
          &       &       & IV-g  & 0.378 & 9.658 \\
          &       &       & IV-TMLE & 0.686 & 2.427 \\ \cline{3-6}
          &       & \multicolumn{1}{l}{$\psi_v$} & TSLS  & 0.416 & 1.277 \\
          &       &       & IV-g  & 0.456 & 1.908 \\
          &       &       & IV-TMLE & 0.875 & 0.822 \\ \cline{2-6}
          & \multicolumn{1}{l}{SL} & \multicolumn{1}{l}{$\psi_c$} & IV-g  & 0.078 & 0.409 \\
          &       &       & IV-TMLE & 0.094 & 0.128 \\ \cline{3-6}
          &       & \multicolumn{1}{l}{$\psi_v$} & IV-g  & 0.107 & 0.212 \\
          &       &       & IV-TMLE & 0.153 & 0.516 \\ \hline\hline
    \end{tabular}%
\end{table}%

\section{Motivating example: the COPERS trial} \label{Sec:COPERS}

We now illustrate the methods in practice by applying each in turn to the motivating example. 
The COping with persistent Pain, Effectiveness Research in Self-management trial (COPERS)  was a randomised controlled  trial across 27 general practices and community services in the UK. It recruited 703 adults with musculoskeletal pain of at least 3 months duration, and randomised 403   participants  to  the active intervention and a further 300 to the control arm. The mean age of participants was 59.9 years, with 81\% white, 67\% female, 23\% employed, 85\% with pain for at least 3 years, and 23\% on strong opioids.

Intervention participants were offered 24 sessions introducing them to cognitive behavioural (CB) approaches designed to promote self-management of chronic back pain.  The sessions were delivered over three days within the same week with a follow-up session 2 weeks later.  At the end of the 3-day course participants received a relaxation CD and self-help booklet. Controls received usual care and the same relaxation CD and self-help booklet. 

The primary outcome was pain-related disability at 12 months, using the Chronic Pain Grade (CPG) disability sub-scale. This is a continuous measure on a scale from 0 to 100, with higher scores indicating worse
pain-related disability.

In the active treatment, only  179  (45\%)  attended all 24 sessions, and 322 (86.1\%) received at least one session.    The control arm participants had no access to the active intervention sessions. Participants and group facilitators  were not masked to the study arm they belonged to. 

The intention-to-treat analysis found no evidence that the COPERS intervention had an effect on 
 improving pain-related disability at 12 months  in people with long-established, chronic musculoskeletal
pain ($-1.0$, $95$\% CI $-4.8$ to $2.7$). 

Poor attendance to the sessions was anticipated, and so obtaining causal treatment effect estimates was a pre-defined objective of the study. 
The original report included a causal treatment effect analysis using TSLS, using a binary indicator for treatment received (attending at least half of the sessions), and 
assuming that randomisation was a valid instrument for treatment received
 \citep{Taylor2016HTA}. The IV model adjusted for the following baseline covariates: site of recruitment, age, gender and HADS score and the  CPG score at baseline.
 This IV analysis found no evidence of  a treatment effect on CPG at 12 months amongst the compliers ( $-1.0$, $95$\% CI $-5.9$ to $3.9$).

The COPERS study also performed a number of subgroup analyses to investigate treatment effect heterogeneity, but did not carry out IV analysis with effect modification. However,  treatment heterogeneity in the causal effect is still of  interest. 

For our re-analyses, the data set consists of  652  participants followed up for 12 months, 374 allocated to active treatment, and  278 in the control (93\% of those recruited).  Thirty-five individuals  (5\%) have  missing primary outcome data, and a further  4 (  <1\%)  have missing baseline depression score, leaving a sample size of 613.

We focus on the causal effect of receiving at least one treatment session as a function of  depression at baseline  measured using the  Hospital Anxiety and Depression Scale (HADS).
 
We argue that random allocation is a valid IV:  the assumptions concerning unconfoundedness and instrument relevance are  justified by design. The exclusion restriction assumption seems plausible with our choices for $A$, as only those participants receiving at least one training sessions would know how to use the CB coping mechanisms and potentially to improve their disability. It is unlikely that that random allocation has a direct effect, though since participants were not blinded to their allocation, we cannot completely rule out  some  psychological effects  of knowing one belongs to the control or active group  on pain and disability. 

We perform each of the methods in turn, TSLS, IV-g and IV-TMLE to estimate $\mbox{ATT}(v)$.
As Table \ref{binexpCOPERS} summarises, the use of DR methods, even after using Super Learner  does not result in a material change in the point estimates or SEs, compared to standard TSLS.  
All five estimators result in the same conclusions, namely that there is no evidence of an average treatment effect in the treated,  and  also that there is no evidence of effect modification by baseline depression. This result could be  due to small numbers of participants in the trial, or indeed our definition of being exposed to treatment (attending at least one session).  Nevertheless, the direction of the treatment effect modification is interesting, indicating that the treatment may benefit more those with higher depression symptoms at  baseline, suggesting a reduction in the disability score.

\begin{table}[h]
  \centering
  \caption{ATT of the COPERS intervention on CPG, with  all-or-nothing binary exposure $A$, 
main effect $\psi_{c}$ and effect modification by depression $\psi_{v}$.}
    \begin{tabular}{lrrrrrr}
\hline
          & \multicolumn{1}{l}{$\psi_{c}$} & \multicolumn{1}{l}{SE} & \multicolumn{1}{c}{95\% CI} & \multicolumn{1}{c}{$\psi_{v}$}& \multicolumn{1}{l}{SE}  & \multicolumn{1}{c}{95\% CI} \\
\hline
  TSLS  & 2.94  & 4.67  &(-6.21, 12.09) & -0.58 & 0.57 & (-1.70, 0.54) \\
    IV-g & 2.78  &4.66 & (-6.35, 11.91)& -0.53 & 0.54 & (-1.59, 0.53) \\
    IV-g SL & 2.10 &  4.75 & (-7.21, 11.41)  & -0.45 &0.54 &  (-1.51, 0.61)\\
    IV-TMLE  & 3.16  & 4.74&(-6.13, 12.45)  &  -0.64  & 0.56 & (-1.74, 0.46) \\
   IV-TMLE SL & 2.22 & 4.88  & (-7.34, 11.78)& -0.51 & 0.58 & (-1.65, 0.63)\\
\hline

    \end{tabular}%
  \label{binexpCOPERS}\end{table}%

\section{Discussion}\label{Sec:discussion}

This paper compared the performance of two doubly robust estimators for the ATT conditional on a baseline covariate, i.e. $\mbox{ATT}(v)$, in the presence of unmeasured confounding, but  where a valid (conditional) IV is available.
These estimators were implemented  with and without the use of data-adaptive estimates of the nuisance parameters.  We have demonstrated empirically through simulations that the IV-g estimator has good finite sample performance when using data-adaptive fits for the nuisance parameters, provided the parametric model assumed for the treatment effect curve is correctly specified.  The IV-TMLE does not rely on  a correctly specified parametric working model, and instead models the whole treatment effect curve, projecting the final estimates onto the working model of interest. This allows us to define the parameters of interest even under a misspecified treatment effect curve. However, it is less efficient compared with the IV g-estimator when the parametric working model for the treatment effect curve is correctly specified. The g-estimator on the other hand can suffer large biases when the assumed treatment effect curve is misspecified.
As the simulations show, the use of data-adaptive fits for the nuisance models greatly reduces bias, and improves coverage for both estimators, resulting in much smaller RMSEs, when compared with using parametric nuisance models, and thus data-adaptive fits should be used. 

The methods were motivated and tested in the context of  estimating the ATT with effect modification in RCTs with non-adherence to randomised treatment with binary exposure and a continuous outcome. 
 However,  the methods presented here  are applicable to other settings. One situation may be where the IV assumptions are  believed to be satisfied only after conditioning on baseline covariates, making this applicable to certain  observational settings. Extensions to situations with continuous exposure are also straight-forward if one is prepared to assume linearity of the treatment effect curve \citep{Toth2016,Vansteelandt2015}. 
 
 We have focused on the $\mbox{ATT}(v)$ as the estimand of interest,  but \citet{Ogburn2015} have shown that the same functional of the observed data can be used to identify under monotonicity the local average treatment effects conditional on baseline covariates,  $\mbox{LATE}(v)$.  
In fact, much of the previous literature regarding estimation of instrumental variable models with covariates has assumed monotonicity. In particular, for the special case where $V=W$,  previous methods include full parametric specifications suitable when both the IV and exposure are binary  \citep{Little1998,Hirano2000} as well as semi-parametric models \citep{Abadie2003}.
 In the case where $V$ is null,  \citet{Frolich2007} characterised two distinct non-parametric estimation methods, while  \citet{Tan2006} proposed a DR estimator which is consistent when the instrument propensity score  and either the outcome or the exposure  parametric models  are correctly specified.

For the  $\mbox{ATT}(v)$,  \citet{Robins1994} proposed DR estimators in settings where $V=W$, while  \citet{Tan2010} did so in settings where  $V$ is a strict subset of $W$ respectively. The DR estimator presented by  \citet{Okui2012} and \citet{Vansteelandt2015}  builds on the work of \citet{Tan2010}. For the special case when $V$ is null,  \citet{Vansteelandt2015} proposed other DR estimators which are locally efficient, and also constructed a bias-reduced DR IV estimator. Several authors have proposed data-adaptive estimators for the  linear IV model with no effect modification, beginning with  a TSLS  where the first stage in fitted using LASSO with a data-adaptive penalty \citep{Belloni2012}. 
The bias-reduced DR IV estimator has also been implemented when $V$ is null using data-adaptive fits for the  conditional mean outcome in the unexposed $\omega(W)$ \citep{Vermeulen2016}.  \citet{Chernozhukov2016} proposed two other IV DR data-adaptive estimators  and gave conditions under which data-adaptive fits can be used for the law of  the instrument $Z$ given $W$, $g(W)$, the treatment model $\pi(Z,W)$ and $\omega(W)$. Comparing these DR estimators to the those presented here would be a promising avenue for future research.  

The present study has some limitations.  Firstly, we did not use sample-splitting in our estimators. Evaluating the effect of doing so in point estimation and variance estimation is a promising extension. In addition, we did not seek to quantify the rates of convergence attained by algorithms included in the SL library. This is because in general the rates of convergence of the individual machine learning algorithms depend on the number of included variables, and other tuning parameters, making the assessment of rates of convergence complex. A potential promising solution  for this could be to include the highly adaptive lasso (HAL) \citep{Benkeser2016} in the SL library, as this has been proven under sufficient regularity conditions to converge at rates faster than $n^{-\frac{1}{4}}$.
 
A number of extensions to the work presented here are of interest. The IV-g method  implemented  here jointly   estimates $\omega(W)$ and $m(W)$, and thus used parametric models for both. This is not necessary, and an alternative strategy where  $\omega(W)$ is  estimated beforehand and  the fitted values are plugged into the estimating equation \eqref{IV-geq} is possible, thus allowing the use of data-adaptive fits for the model $\omega(W)$.  Future work could extend the bias-reduced DR estimator to the linear IV model with effect modification, and compare this with IV-TMLE and a fully data-adaptive version of the IV-g estimator.

\section*{Acknowledgements}
We thank Prof. Stephanie Taylor and the COPERS trial team  for access to the data. We also thank Prof. Stijn Vansteelandt for commenting on an earlier draft of this paper, and Boriska T\'oth  for sharing her code implementing IV-TMLE.\\
Karla DiazOrdaz was supported by UK  Medical Research Council Career development award in Biostatistics MR/L011964/1 and UK Wellcome Trust Institutional Strategic Support Fund-- LSHTM Fellowship 204928/Z/16/Z.

\section*{Appendix}
\subsection{Consistent data-adaptive g-estimation}

Here, we give here a sketch of the proof of consistency of the data-adaptive g-estimator.  Similarly to \citet{Chernozhukov2016} and \citet{Farrell2015}, we let $\mP$ be the class of probability distributions for $O$ that obey the partially linear IV model, such that for each $P\in\mP$, the 
restrictions  $E[Y - Am_0(W) | W] =\omega_0(W)$,  $E[A\vert Z,W]=\pi_0(Z,W)$, and  $E[Z\vert W]=g_0(W)$ hold. Let $\eta_0$ denote the nuisance functional describing $g_0(W)$, $\omega_0(W)$, and $\pi_0(Z,W)$. For simplicity, we sketch the arguments under the null, that is $m_0(W;\psi_0)=0$. In addition, unlike \citet{Chernozhukov2016}, we do not use sample splitting, and proceed instead under empirical processes conditions which are from now on assumed to hold. 

Denote by $\epsilon_Y=Y- \omega_0(W)$ and $\epsilon_A=\pi_0(Z,W)- E(\pi_0(Z,W)\vert W)$. Since the instrument $Z$ is binary, we can write $E(\pi_0(Z,W)\vert W)=\sum_{Z} \pi_0(Z,W)g_0(W)$, where we use the shorthand $g_0(W)=g_0(Z=z,W)=Pr(Z=z\vert W)$. 

We want to find conditions  guaranteeing that $\sqrt{n}\left(\widehat{\psi}-\psi_0\right)=o_p(1)$, where $\widehat{\psi}$ has been estimated with the IV g-estimator which used data-adaptive estimates for the nuisance parameters $\eta$.

We begin by writing 
\begin{eqnarray}\label{decomposition}
&&\sqrt{n}\left(\widehat{\psi}-\psi_0\right)=\\
\nonumber&&\quad\frac{1}{\sqrt{n}}\sum_i\mS(O_i;\psi_0,\widehat{\eta_0}) -E_p[\mS(O_i \psi_0, \widehat{\eta_0})]-\mS(O_i,\psi_0,\eta_0)-E_p[\mS(O_i \psi_0, \eta_0)] +  \sqrt{n} E_{P}[\mS(O_i,\psi_0,\widehat{\eta_0})],
\end{eqnarray}
where $\mS$ is the score corresponding to the estimating equation \eqref{IV-geq}, with $\sigma_0^2=1$, i.e.:  
\begin{equation}\mS(O_i; \psi_0, \eta_0)=\left\{\pi_0(Z_i,W_i)-E_p[\pi_0(Z_i,W_i)g_0(W_i)]\right\}\left\{Y_i-\omega_0(W_i)-m_0(W_i;\psi_0)A_i\right\},
\end{equation}

The first part can be shown to be $o_P(1)$  where  $\norm{\mS(O_i,\psi_0,\widehat{\eta_0}) - \mS(O_i,\psi_0,{\eta_0})}=o_P(1)$, by Chebyshev's inequality.

Therefore, we want to give sufficient conditions for
\begin{equation}\label{Asump1}\norm{\mS(O_i,\psi_0,\widehat{\eta_0}) - \mS(O_i,\psi_0,{\eta_0})}_p=o_P(1).\end{equation}  

Using  recursive expansion  around each true nuisance functional,  $\mS(O_i,\psi_0,\widehat{\eta_0})$ can be written as 
\begin{eqnarray*}
&=& \{\pi_0(Z,W) - \sum_{z}{\pi_0}(Z,W)g_0(W)\}  \{Y-{\omega_0}(W)\} +  \{\pi_0(Z,W) - \sum_{z}{\pi_0}(Z,W)g_0(W)\} \left(\omega_0(W)-\widehat{\omega_0}(W)\right)\\
&+& \sum_{z}{\pi_0}(Z,W)\left(g_0(W)-\widehat{g_0}(W)\right) \{Y-{\omega_0}(W)\}  +\sum_{z}{\pi_0}(Z,W)\left(g_0(W)-\widehat{g_0}(W)\right) \left(\omega_0(W)-\widehat{\omega_0}(W)\right)\ \\
&+& \{ \left(\pi_0(Z,W) - \widehat{\pi_0}(Z,W)\right)  - \sum_z \left( \pi_0(Z,W) - \widehat{\pi_0}(Z,W)\right) g_0(W)\}  \{Y-{\omega_0}(W)\} \\
&+&  \{ \left(\pi_0(Z,W) - \widehat{\pi_0}(Z,W)\right)  - \sum_z \left( \pi_0(Z,W) - \widehat{\pi_0}(Z,W)\right) g_0(W)\}  \left(\omega_0(W)-\widehat{\omega_0}(W)\right)\  \\ 
&+&  \sum_z \left(  \widehat{\pi_0}(Z,W)- \pi_0(Z,W) \right) \left(g_0(W)-\widehat{g_0}(W)\right) \{Y-{\omega_0}(W)\} \\
&+& \sum_z \left(  \widehat{\pi_0}(Z,W)- \pi_0(Z,W) \right) \left(g_0(W)-\widehat{g_0}(W)\right) \left(\omega_0(W)-\widehat{\omega_0}(W)\right),
\end{eqnarray*} 
which can be further simplified to:
\begin{eqnarray*}
&=&  \epsilon_A\epsilon_y + \epsilon_A 
\{\omega_0(W)-\widehat{\omega_0}(W)\} 
- \sum_z \pi_0(Z,W) \left\{g_0(W)-\widehat{g_0}(W)\right\} \epsilon_y\\
&-& \sum_z\pi_0(Z,W) \left(g_0(W)-\widehat{g_0}(W)\right) \left(\omega_0(W)-\widehat{\omega_0}(W)\right)\\
&+& \epsilon_y \left\{\left(\widehat{\pi_0}(Z,W) -\pi_0(Z,W)\right) - \sum_z \left(\widehat{\pi_0}(Z,W) -\pi_0(Z,W)\right)g_0(W)\right\} 
\\&+& \left(  \widehat{\pi_0}(Z,W)- \pi_0(Z,W) \right) \left(\omega_0(W)-\widehat{\omega_0}(W)\right)  \\
&-&   \sum_{z}\left(  \widehat{\pi_0}(Z,W)- \pi_0(Z,W) \right) g_0(W) \left(\omega_0(W)-\widehat{\omega_0}(W)\right)\\
 &+&
\epsilon_y \sum_{z}\left(  \widehat{\pi_0}(Z,W)- \pi_0(Z,W) \right) \left(g_0(W)-\widehat{g_0}(W)\right)\\
&-&  \sum_{z}\left(  \widehat{\pi_0}(Z,W)- \pi_0(Z,W) \right)\left(g_0(W)-\widehat{g_0}(W)\right)\left(\omega_0(W)-\widehat{\omega_0}(W)\right).
\end{eqnarray*} 
Therefore 
\begin{eqnarray*}
&&\norm{\mS(O_i,\psi_0,\widehat{\eta_0}) - \mS(O_i,\psi_0,{\eta_0})}_p\\
&\leq &\norm{\epsilon_A \{\omega_0(W)-\widehat{\omega_0}(W)\} }_P 
 + \norm{\epsilon_y \left\{\left(\widehat{\pi_0}(Z,W) -\pi_0(Z,W)\right) - \sum_z \left(\widehat{\pi_0}(Z,W) -\pi_0(Z,W)\right)g_0(W)\right\} }_P
\\&+& \norm{\epsilon_y \sum_{z}\pi_0(Z,W)\left( \widehat{g_0}(W)-g_0(W) \right) }_P \\
&+& \norm{\left(\omega_0(W)-\widehat{\omega_0}(W)\right) \left\{\left(\widehat{\pi_0}(Z,W) -\pi_0(Z,W)\right) - \sum_z \left(\widehat{\pi_0}(Z,W) -\pi_0(Z,W)\right)g_0(W)\right\} }_P
\\&+& \norm{\left(\omega_0(W)-\widehat{\omega_0}(W)\right) \sum_{z}\pi_0(Z,W)\left( \widehat{g_0}(W)-g_0(W) \right) }_P\\
 &+& \norm{\sum_z \left(  \widehat{\pi_0}(Z,W)- \pi_0(Z,W) \right) \left(g_0(W)-\widehat{g_0}(W)\right) \left(\omega_0(W)-\widehat{\omega_0}(W)\right)}_P\\
&\leq &
\sqrt{C} \norm{\omega_0(W)-\widehat{\omega_0}(W)}_P  + 
\sqrt{C}\norm{\left\{\left(\widehat{\pi_0}(Z,W) -\pi_0(Z,W)\right) - \sum_z \left(\widehat{\pi_0}(Z,W) -\pi_0(Z,W)\right)g_0(W)\right\} }_P
\\&+& \sqrt{C}\norm{\left(\pi_0(1,W)-\pi_0(0,W)\right)\left( \widehat{g_0}(W)-g_0(W) \right) }_P
\\&+& \norm{\left(\omega_0(W)-\widehat{\omega_0}(W)\right)\left\{\left(\widehat{\pi_0}(Z,W) -\pi_0(Z,W)\right) - \sum_z \left(\widehat{\pi_0}(Z,W) -\pi_0(Z,W)\right)g_0(W)\right\} }_P
\\&+& \norm{\left(\omega_0(W)-\widehat{\omega_0}(W)\right)\left(\pi_0(1,W)-\pi_0(0,W)\right)\left( \widehat{g_0}(W)-g_0(W) \right) }_P\\
&+& \norm{\sum_z \left(  \widehat{\pi_0}(Z,W)- \pi_0(Z,W) \right) \left(g_0(W)-\widehat{g_0}(W)\right) \left(\omega_0(W)-\widehat{\omega_0}(W)\right)}_P,
\end{eqnarray*} 
where we assume that there exists a  constant $C> 0$, such that
\begin{eqnarray}\label{boundedres}
\nonumber P(E[\{\pi_0(Z,W)- \sum_z\pi_0(Z,W)g_0(W)\}^2]\leq C)&=&1, \ \text{ and}\\
P(E[\left\{Y- \omega_0(W) \right\}^2]\leq C)&=&1.
\end{eqnarray} 
Finally, since $\norm{\left({\pi_0}(1,W)- {\pi_0}(0,W)\right)}$ is bounded and
\begin{eqnarray}\label{conds1} 
\norm{\widehat{\pi_0}(Z,W)- \pi_0(Z,W)} &=& o_{P_0}({1}),\\
\norm{ \widehat{\omega_0}(W)- \omega_0(W)}&=&  o_{P_0}({1}),\\
\norm{ \widehat{g_0}(W)- g_0(W)} &=&  o_{P_0}({1}),
\end{eqnarray}
by definition of $\omega_0(W)$ and $g_0(W)$, we conclude that the assumption \eqref{Asump1} needed for the first part of eq. \eqref{decomposition} to be  $o_{P}({1})$ holds. 

Now, for  the second term in eq. \eqref{decomposition},  we want conditions such that 
\begin{equation}
\label{secondpart}
\sqrt{n} E_{P}[\mS(\psi_0,\widehat{\eta_0})]=o_{P}({1}),\end{equation}
we have 
\begin{eqnarray*}
\sqrt{n} E_{P}[\mS(\psi_0,\widehat{\eta_0})] &=& \sqrt{n} E_{P}[\widehat{\pi_0}(Z,W) - E_P[\widehat{\pi_0}(Z,W)](Y-\widehat{\omega_0}(W))]\\
&=& \sqrt{n} E_{P}[\widehat{\pi_0}(Z,W) - \sum_z\left\{\widehat{\pi_0}(Z,W)\widehat{g_0}(W)\right\}(Y-\widehat{\omega_0}(W))]\\
&=& \sqrt{n} E_{P}[\widehat{\pi_0}(Z,W) - \sum_z\left\{\widehat{\pi_0}(Z,W)\widehat{g_0}(W)\right\}(\omega_0-\widehat{\omega_0}(W))]\\
&=& \sqrt{n} E_{P}\left[\widehat{\pi_0}(Z,W) - \sum_z\left\{\widehat{\pi_0}(Z,W)\left(g_0(W)-\widehat{g_0}(W)\right)\right\}\left(\omega_0-\widehat{\omega_0}(W)\right)\right]\\
&=&\sum_z \{\widehat{\pi_0}(Z,W)\left( g_0(W)-\widehat{g_0}(W)\right)\}
\{\omega_0(W)-\widehat{\omega_0}(W)\}.
\end{eqnarray*} 
Now the norm of the first term of this expression is such that
\begin{eqnarray*}
\norm{\sum_z \widehat{\pi_0}(Z,W)\left(g_0(W)-\widehat{g_0}(W)\right)}
&=&
\norm{\widehat{\pi_0}(1,W)\left(g_0(1,W) -\widehat{g_0}(1,W)\right) + \widehat{\pi_0}(0,W)\left(g_0(0,W) -\widehat{g_0}(0,W)\right)}\\
&=&
\norm{\left(\widehat{\pi_0}(1,W)- \widehat{\pi_0}(0,W)\right) \left(g_0(1,W) -\widehat{g_0}(1,W)\right)}\\
&\leq& 
\norm{\left(\widehat{\pi_0}(1,W)- \widehat{\pi_0}(0,W)\right)}\norm{ \left(g_0(1,W) -\widehat{g_0}(1,W)\right)},
\end{eqnarray*} 
where we have used Cauchy-Schwarz inequality in the last step.
Now, since $\norm{\left(\widehat{\pi_0}(1,W)- \widehat{\pi_0}(0,W)\right)}$ is bounded, assuming 
\begin{equation}\label{rate} \norm{ g_0(W)-\widehat{g_0}(W)}\norm{\omega_0(W)-\widehat{\omega_0}(W)}=o_{P}(n^{\frac{-1}{2} }),\end{equation} is sufficient to guarantee eq. \eqref{secondpart} holds.

In summary, to guarantee the data-adaptive IV g-estimator is CAN assumptions \eqref{boundedres}, \eqref{conds1}  and \eqref{rate} need to hold. These conditions are essentially the same found by \citet{Chernozhukov2016}.

\clearpage
\subsection{Extra results}
\begin{table}[h]
  \centering
  \caption{Scenarios excluded from the Bias and Coverage Figures.}
    \begin{tabular}{rlllrrrr} \hline\hline
   \multicolumn{1}{l}{Scenario} & m model & Parameter & Method & \multicolumn{1}{l}{bias} & \multicolumn{2}{c}{MCE CI} & \multicolumn{1}{l}{coverage} \\ \hline
    \multicolumn{1}{l}{$n=500$} \\  \hline
 \multicolumn{1}{l}{$\pi$ mis, $\omega$ cor} & cor   & $\psi_v$ & IV-g$^\text{a}$ & -4.057 & -12.691 & 4.577 & 0.986 \\
                & mis   & $\psi_c$ & TSLS  & -6.881 & -16.826 & 3.064 & 0.683 \\
                & mis   & $\psi_c$ & IV-g$^\text{a}$& -31.033 & -58.624 & -3.442 & 0.882 \\
                 & mis   & $\psi_v$ & TSLS  & 6.785 & -3.348 & 16.918 & 0.998 \\
                 & mis   & $\psi_v$ & IV-g$^\text{a}$ & -45.811 & -144.787 & 53.165 & 0.982 \\
           \multicolumn{1}{l}{$\pi$ mis, $\omega$ mis} & cor   & $\psi_v$ & TSLS  & -2.171 & -6.834 & 2.492 & 0.984 \\
                 & mis   & $\psi_c$ & TSLS  & -6.740 & -16.330 & 2.850 & 0.694 \\
                 & mis   & $\psi_c$ & IV-g$^\text{a}$ & -30.361 & -56.392 & -4.330 & 0.879 \\
                 & mis   & $\psi_v$ & TSLS  & 5.751 & -3.557 & 15.059 & 0.996 \\
                 & mis   & $\psi_v$ & IV-g$^\text{a}$ & -41.902 & -134.385 & 50.581 & 0.984 \\ \hline
    \multicolumn{1}{l}{$n=10,000$} \\  \hline
 \multicolumn{1}{l}{$\pi$ mis, $\omega$ cor} & mis   & $\psi_c$ & TSLS  & -10.101 & -10.169 & -10.032 & 0.000 \\
                 & mis   & $\psi_c$ & IV-g$^\text{a}$ & -9.521 & -9.622 & -9.420 & 0.001 \\
           \multicolumn{1}{l}{$\pi$ mis, $\omega$ mis} & mis   & $\psi_c$ & TSLS  & -10.097 & -10.166 & -10.028 & 0.000 \\
                 & mis   & $\psi_c$ & IV-g$^\text{a}$  & -9.518 & -9.620 & -9.416 & 0.001 \\
                 & mis   & $\psi_c$ & IV-TMLE$^\text{a}$ & -2.333 & -2.374 & -2.291 & 1.000 \\ \hline

			\multicolumn{8}{l}{$^\text{a}$ \footnotesize{All nuisance models fitted parametrically.}} \\
		\hline\hline     \end{tabular}%
  \label{tab:excluded}%
\end{table}%

\clearpage
\bibliographystyle{apalike}

\end{document}